\documentclass[letterpaper,12pt]{article}
\usepackage{osajnl2}

\usepackage{amsmath,amssymb}
\usepackage{subfigure}

\def\mf{\mathbf}

\newcommand{\refeq}[1]{Eq.  (\ref{#1})} 
   
\newcommand{\reffig}[1]{Fig. \ref{#1}}
\newcommand{\leftsub}[2]{{\vphantom{#2}}_{#1}{#2}}
\newcommand{\nCk}[2]{\ensuremath{\leftsub{#1}{C}_{#2}}}

\begin{document}

\title{Computer vision applications for coronagraphic optical alignment and image processing} 

\author{Dmitry Savransky$^{1,*}$, Sandrine J. Thomas$^2$, Lisa A. Poyneer$^1$, Bruce A. Macintosh$^1$ }
\address{$^1$Lawrence Livermore National Laboratory,\\7000 East Ave., Livermore, CA 94550, USA}
\address{$^2$Gemini Observatory,\\670 N. A'ohoku Place, Hilo, Hawaii, 96720, USA}
\address{$^*$Corresponding author: savransky1@llnl.gov}

\begin{abstract}
Modern coronagraphic systems require very precise alignment between optical components and can benefit greatly from automated image processing.  We discuss three techniques commonly employed in the fields of computer vision and image analysis as applied to the Gemini Planet Imager, a new facility instrument for the Gemini South observatory.  We describe how feature extraction and clustering methods can be used to aid in automated system alignment tasks, and also present a search algorithm for finding regular features in science images used for calibration and data processing.  Along with discussions of each technique, we present our specific implementation and show results of each one in operation.
\end{abstract}

\section{Introduction}
The study of extrasolar planets is one of the most exciting areas in modern astrophysics \cite{seager2011exoplanets}.  A new and particularly important technique is direct imaging---blocking the light from a star to allow a nearby planet to be seen \cite{marois2008direct}. This is very challenging as even the brightest extrasolar planets will be $10^4$ to $10^7$ times fainter than their parent star. To successfully image an extrasolar planet, an instrument must be specifically designed to control light scattered by both diffraction and wavefront errors. An example of such an instrument is the Gemini Planet Imager (GPI). GPI  is a new facility instrument for the Gemini South Observatory that is designed to directly detect and spectrally characterize young, Jovian and super-Jovian extrasolar planets, which are difficult to detect with other planet-finding techniques.  Scheduled to go on-sky in 2013, GPI is expected to achieve deeper contrast levels than any instrument currently in operation.  As such, it has very tight alignment tolerances, which are maintained via multiple open and closed loop schemes \cite{macintosh2008gemini}.  To this end, GPI contains a large number of moveable surfaces, as well as multiple internal imaging systems in addition to the main science instrument.  In order to achieve rapid, precise and repeatable internal system alignment, we have automated the processing of this imaging data, utilizing numerous techniques commonly employed in the field of computer vision.  In this paper we discuss three of these applications, and detail their implementation as it applies to GPI components.  

GPI's design has been extensively discussed in previous literature (see, e.g., \cite{macintosh2012gemini}).  Here, we will present a broad overview and focus only on those components and functions relevant to the applications being presented, and common to multiple next-generation, coronagraphic instruments.  GPI can be described in terms of four main subsystems: 
\begin{enumerate}
\item The Adaptive Optics (AO) system, which corrects for atmospheric turbulence \cite{poyneer2004spatially}.
\item The calibration system (CAL)---a modified Mach-Zehnder interferometer, which is responsible for correcting for non-common path errors\cite{wallace2008post}.
\item The science instrument---an integral field spectrograph (IFS) capable of producing dispersed high-resolution spectral images that can then be reconstructed into three dimensional data cubes of two spatial and one wavelength dimensions \cite{maire2010data}.
\item The diffraction control system, which creates high contrast regions in the science image.  GPI's primary diffraction control is achieved with an apodized-pupil lyot coronograph (APLC) \cite{soummer2004apodized}. 
\end{enumerate}

The APLC design addresses several issues associated with applying classical coronagraphs to ground-based telescopes with central obscurations and allows us to remove much of the light diffracted into the halo of the telescope's point spread function (PSF), achieving contrasts of up to $10^{7}$ over the YHJK bands.  The classical Lyot coronagraph employs a focal plane mask (FPM) to block the light of the bright, on-axis, target star. Some of this light is diffracted about the mask, and appears as a series of bright rings at the edge of the pupil when reimaged in the next pupil plane. A Lyot stop placed in this output pupil, known as the Lyot plane, shrinks the pupil and blocks this light, generating regions in the focal plane with very little on-axis light, but which retain the majority of any off-axis source. In the pupil plane, the effect of the focal plane mask is to subtract a version of the input pupil convolved and blurred by the Fourier transform of the focal plane stop. This over-subtracts light outside the pupil and under-subtracts light near the edges (see \cite{sivaramakrishnan2008speckle} for further discussion).

Unfortunately, this process is not perfect. Although the diffracted light is concentrated at the edges of the Lyot pupil, ringing from the hard-edge, focal-plane stop produces some residual light throughout the entire pupil; reducing this requires stopping the pupil down by a very large amount, producing even more diffraction as well as significantly impacting the amount of off-axis light transmitted. The presence of a central obscuration such as a Cassegrain secondary mirror, and the `spiders' that support it, also diffracts more light into the halo and similarly decreases the efficiency of the system, greatly lowering the achieved contrast. By introducing an apodizer into the pupil plane, carefully matched to the dimensions of the focal plane mask, the residual diffraction can be channeled almost completely into a sharp bright ring outside the input pupil and almost completely rejected without undersizing the Lyot stop (see \cite{soummer2004apodized} for discussion). 

For this to work, the star must be precisely ($< 0.1  \, \lambda/D$, where $\lambda$ is the wavelength and $D$ the telescope diameter) centered on the focal plane mask.  For comparison, in coronagraphic imaging at the Keck Observatory, the point spread function moves randomly on the FPM by up to 1 $\lambda/D$ [Marois, 2012; personal communication].  As the size of the PSF is wavelength dependent (scaling linearly with wavelength), the size of the FPM is determined by the band in which the instrument operates, and the design of the apodizer must be matched to a particular FPM.  This makes the APLC chromatic, and necessitates a separate FPM and apodizer for each wavelength band in which we wish to operate. Manufacturing the apodizers is also challenging; GPI uses metal-on-glass microdot masks that have some further inherent chromaticity \cite{sivaramakrishnan2008astrometry}, requiring separate focal plane masks for each ~20\% bandpass. Residual diffraction from obscurations such as the spiders will result in narrow bright features in the output Lyot plane that must be also be blocked. This leads to tight alignment tolerances of less than 1\% of the pupil diameter. These can be relaxed by oversizing the blocked area of the masks, but at a cost of sensitivity and throughput.  The Lyot stop can also be used to block other bright features, such as those due to non-functioning actuators in the MEMS, which would otherwise spread light everywhere in the focal plane.

\reffig{fig:gpi_schematic} shows a schematic view of the light path through GPI.  Light entering GPI from the telescope first passes through the AO system, whose moveable components include an input fold mirror, two deformable mirrors (DMs)---a piezo-electric and a MEMS mirror in a woofer-tweeter arrangement, and a tip-tilt stage on the woofer.  These components operate in closed loop with a Shack-Hartmann wavefront sensor (WFS).  The beam is split via a dichroic with the visible light portion steered onto the WFS with a pair of pointing and centering mirrors, while the IR  light passes through one of the pupil plane apodizers (held in a filter wheel) and into the CAL system.  The first focal plane in the CAL system contains the filter wheel holding all of the FPMs.  These are reflective optics with small central holes that allow on-axis light to pass through while off-axis light is reflected.  A portion or the on-axis light is sent to the low-order wavefront sensor (LOWFS)---a near-IR Shack-Hartmann.  The remaining on-axis light is spatially filtered to generate a spherical reference wavefront, interfered with a portion of the off-axis light, and imaged onto an IR detector.  This creates a high-order wavefront sensor (HOWFS), sensitive to all spatial frequencies not blocked by the FPM.    Reconstructed LOWFS and HOWFS wavefronts are applied as reference centroid offsets to the main AO loop at 1 Hz, correcting for non-common path errors and maintaining a stable wavefront.  The HOWFS detector can also be used as a pupil viewer by only imaging light from one leg of the interferometer.

A second set of pointing and centering mirrors then centers the beam on the IFS pupil, which contains a third filter wheel with the Lyot stops.  In addition to the Lyot stop prescribed by the APLC design, extra stops are included with oversized central obscurations and spiders to simplify alignment in cases where some throughput can be sacrificed, and to ensure that a dark hole can be created even if diffraction about the telescope spiders is more significant than expected.  Any light diffracted in this plane will be spread throughout the science image, significantly decreasing achieved contrast, so it is important to ensure that all light can be blocked in any circumstances.  The stops also contain tabs attached to the spiders used to block light  from bad actuators in the MEMS DM.  The IFS also includes a deployable pupil viewing camera that can be used for imaging the IFS entrance pupil, just as the HOWFS detector can be used to image the CAL pupil, and the AO WFS can be used to track the telescope pupil.  It should be noted that the IFS pupil viewer is not a science-grade instrument, does not have an absolute photometric calibration, and suffers from gain drifts and high noise.  These three imaging systems, along with the science instrument itself, provide all of the information required to align the system.

\reffig{fig:pupil_schematic} shows a schematic view of the image at the IFS pupil, demonstrating that each of the APLC components, along with the telescope pupil, must be closely aligned for the system to function.  The MEMS DM plane is typically used as a common reference, with each of the components aligned to this plane independently of the others.  Table \ref{tbl:tolerances} contains the alignment tolerances of the various components, calculated based on the APLC design, and numerical modeling of the system with a target contrast of $10^{7}$. Alignment of all of the optics up to the apodizer is done with data from the AO WFS.   The sensor measures tip and tilt, providing pointing information for the tip/tilt and AO pointing and centering mirrors.  To center the telescope pupil, the wavefront sensor intensities are autocorrelated to find any asymmetries, which are then removed by actuating the input fold mirror.  The AO WFS is aligned to the MEMS plane by putting a known shape (a pattern of `poked' actuators) on the DM,  and then cross-correlating the input shape with the measured AO WFS phase \cite{stone2001fast}.  The measured offsets are corrected with the AO pointing and centering mirrors.

Mis-alignment of both the apodizer and Lyot stop from the MEMS plane can be measured in imaging data from the pupil-viewer or HOWFS detector.  As phases on the DMs map to intensities in the Lyot plane, it is relatively simple to identify individual MEMS pokes, especially when the pupil viewer is slightly defocused.  Locating the apodizer and Lyot stop, however, requires us to automatically identify features on the masks.  Section \ref{sec:ellipse_finding} details the application of the generalized Hough transform for identifying mask contours, while \S\ref{sec:multiline_finding} describes a clustering technique that is useful in separating pixels belonging to different features of interest (such as the Lyot mask spiders).  Finally, while GPI is specifically designed to block the light from the target star so that light from the (much dimmer) planets can be observed, it is still important to be able to locate the target precisely in the final science images.  To accomplish this, a grid is printed on the apodizer mask, producing four satellite spots at equal distances from the true image center in the focal plane \cite{sivaramakrishnan2008astrometry,marois2008accurate}.  This distance scales with wavelength, and the specific location of the spots can change with instrument alignment.  Furthermore, the specific location of the spots and their magnitudes are an important diagnostic tool, so it is very useful to have an automated method for finding them, using as few prior assumptions as to their location as possible.  Section \ref{sec:satfinding} describes a search algorithm that accomplishes this and is generally applicable for efficient location of vertices of regular geometric shapes.  This same technique is also be applied to finding MEMS pokes in pupil viewer images.

\section{Ellipse Finding}\label{sec:ellipse_finding}
Many components of GPI, including the apodizers, the Lyot stop and the telescope pupil itself, form circular shapes when imaged at a pupil plane.  As various misalignment, projection, and focus effects can cause these circular shapes to become slightly deformed, we wish to generalize the automated location of these shapes as fitting ellipses to the data and do so by applying a Hough transform.  The Hough transform is a standard tool in computer vision for detecting both analytical curves and arbitrary shapes and is widely used in various image processing applications \cite{ballard1981generalizing}.  Recent work on the Hough transform applied specifically to ellipses has yielded a highly memory-efficient algorithm \cite{xie2002new}.  

The Hough transform seeks to find a parameter set that maximizes the fit of a set of pixels to a specific class of shapes.  For an arbitrary analytic curve of the form $f(\mf x, \mf p)  = 0$, where $\mf x$ is the image (pixel) location and $\mf p$ is an arbitrarily sized vector of parameters, this is equivalent to solving the constraint equation:
\begin{equation}\label{eq:gradient}
\left.\frac{\mathrm{d}f}{\mathrm{d}\mf x}\right\vert_{\mf x, \mf p} = \mf 0 \,.
\end{equation}
The algorithm involves the formation of an accumulator array, $A(\mf p)$.  For each pixel $\mf x$, all parameter vectors satisfying definition of the curve and \refeq{eq:gradient} are found and the corresponding locations in the accumulator array are incremented.  At the end of this process, local maxima in $A$ will correspond to curves of $f$ in the original image. \cite{ballard1981generalizing}

Following the formalism of \cite{xie2002new}, an ellipse may be defined via four parameters:
\begin{equation}
\frac{(x - x_0)^2}{a^2} + \frac{(y - y_0)^2}{b^2} = 1 \,,
\end{equation}
where $(x_0,y_0)$ is the ellipse center and $a,b$ are the semi-major and semi-minor axes, respectively.  A fifth parameter, $\theta$, representing the rotation of the ellipse axes with respect to the image axes is added to fully generalize the description.  Thus, if we assume that an arbitrary pair of pixels, $(x_1,y_1)$ and $(x_2,y_2)$, form the endpoints of the ellipse major axis, the five parameters are given by:
\begin{align}
x_0 &= (x_1 + x_2)/2\\
y_0 &= (y_1 + y_2)/2\\
a &= \sqrt{(x_2 - x_1)^2 + (y_2 - y_1)^2}/2\\
\theta &= \tan^{-1}\left((x_2 - x_1)/ (y_2 - y_1)\right) \,.
\end{align}
Given any other point $(x_3,y_3)$ on the ellipse, and defining:
\begin{align}
d &\triangleq \sqrt{(x_3 - x_0)^2 + (y_3 - y_0)^2} \\
f &\triangleq \sqrt{(x_3 - x_2)^2 + (y_3 - y_2)^2} \,,
\end{align}
the semi-minor axis is approximated as:
\begin{equation}
b = \sqrt{\frac{a^2\left[(2 a d)^2 - (a^2 + d^2 - f^2)^2\right]}{a^4 -  (a^2 + d^2 - f^2)^2}} \,.
\end{equation}
This allows us to define the accumulator array only for the semi-minor axis.  The maxima of the accumulator array will represent candidate ellipse semi-minor axes (along with the parameters used to calculate them).  The overall maximum can then be taken as the most likely ellipse, or all candidates whose accumulator values exceed a pre-defined threshold can be retained.

For GPI, this algorithm is applied to pupil viewer images in order to align the apodizers and Lyot masks to the MEMS plane.  The pupil viewer produces 320x240 pixel images, capturing the IFS entrance pupil in a square approximately 232 pixels in height.  These require some pre-processing before the main algorithm is applied.  Starting with a dark subtracted image, a $3\times3$ pixel median filter is applied to remove isolated hot and cold pixels and smooth the image.  Next, in order to decrease the processing time for the Hough transform, a set of candidate pixels is identified by finding pixel locations where the difference between values of neighboring pixels is greater than a threshold set by the image median value.  That is, for an image represented by and $m\times n$ matrix $M$, the set of candidate pixels is defined as:
\begin{equation}
\left\{i,j : \vert M_{i-1,j}-M_{i+1,j} \vert > T\right\} \cup \left\{i,j : \vert M_{i,j-1}-M_{i,j+1} \vert > T\right\} \,, i,j \in [1,m] \times [1,n]  \,,
\end{equation}
where $T$ is the threshold, set to a multiple of the image median.  The Hough transform is then applied to only these candidate pixels, and the global maximum of the accumulator array is retained as the best-fit solution.  \reffig{fig:pupil_hough} illustrates these steps on a sample data set imaging one of GPI's apodizers. Note that it is not necessary for the candidate set to include all of the pixels belonging to the best-fit ellipse, or even an entire closed curve.  The algorithm works when as few as 50\% of the pixels belonging to the feature of interest are identified.

Once the center of the apodizer (or Lyot mask) is found, it is necessary to also find the center of the DM.  To do so, we apply a symmetric shape to the DM in the form of four `poked' actuators, producing four symmetric bright spots in the pupil plane.  The points are located using a difference image between the poked and unpoked images and by cross-correlating with a template, or by applying the algorithm described in \S\ref{sec:satfinding}.  The intersection of these four points is compared to the center of the best-fit ellipse and the difference is used to adjust the beam pointing and align the apodizer and Lyot stop planes to the MEMS DM plane, as illustrated in \reffig{fig:pupalign}.  

The computation required by the Hough transform is relatively minimal, with all processing performed on a single, conventional CPU.  The execution time of these alignments is dominated by the camera integration and readout times and communication overhead associated with sending mirror commands to the MEMS DM.  This algorithm is also more robust than more conventional `center of gravity' image localization techniques, which are much more sensitive to image noise, illumination variations and flat-field errors, and whose results are highly dependent on which pixels are included in the processing.  At the same time, illumination changes in the system and bias drifts in the pupil viewing camera make it difficult to generate a stable template to use for matched filtering or other correlation techniques, whereas the Hough transform approach will work as long as there is a measurable intensity difference between the central obscuration and the illuminated portion of the pupil.

\section{Multiple Line Finding}\label{sec:multiline_finding}
An alternative approach to the pupil plane alignment algorithm described in \S\ref{sec:ellipse_finding} is to use the HOWFS rather than the pupil viewer.  This has the benefit of a faster readout time, and can be used at the same time as the science instrument, without having to deploy the pupil viewer.  Unfortunately, the HOWFS is lower resolution than the pupil viewer, capturing the masks in only about $40\times40$ pixel squares, and individual poked actuators are not as easy to locate as in the pupil images, especially at lower light levels.  The solutions found from poke images on the HOWFS are therefore not as precise as those from the pupil viewer data.  To address this, we instead can use whole rows and columns of MEMS actuators to create a cross pattern on the HOWFS.  The pixels belonging to this shape are much brighter than everything else in the image (after hot pixels have been removed) and are therefore simple to extract.  The question then becomes, however, how to automatically group the resulting pixels into four separate line segments.

To accomplish this, we use $k$-mean clustering---a commonly applied algorithm in computer vision used to partition a set of $n$ observations into $k$ topographically distinct clusters. \cite{macqueen1967some}  Given a set of vectors $\{\mf r_i\}_{i=1}^n = \{\mf r_1, \mf r_2,\ldots, \mf r_n\}$ (in this case, the vectors are 2D pixel locations), this algorithm seeks to find the subsets $\{S_i\}_{i=1}^k$ such that:
\begin{equation}
\arg\min_{\{S_i\}} \sum_{i=1}^k\sum_{\mf r_k \in S_i} \Vert r_k - \boldsymbol\mu_i \Vert^2 \,,
\end{equation}
where $\mu_i$ is the mean of set $S_i$.  This is achieved iteratively: an initial set of $k$ means,  $\{ \boldsymbol\mu_i\}_{i=1}^k$, is randomly generated.  The subsets are then assigned as:
\begin{equation}
S_i = \left\{ \mf r_k : \Vert r_k - \boldsymbol\mu_i \Vert^2 \le \Vert r_k - \boldsymbol\mu_j \Vert^2 \,, \forall i \ne j \right\} \,,
\end{equation}
and the cluster means are subsequently updated based on the new sets:
\begin{equation}
\boldsymbol\mu_i = \frac{1}{n_i} \sum_{\mf r_k \in S_i} \mf r_k \,,
\end{equation}
where $n_i$ is the current size of set $S_i$.  These two steps are repeated until no further changes occur in the subset assignments.

As we have only four clusters in our application, and we can intelligently choose initial means by placing them in the center of the four quadrants of the HOWFS image, we can typically achieve convergence in only 2 to 3 iterations.  Once the four line segments have been identified, we fit lines to each one, and find the intersection point as before.  This procedure is illustrated in \reffig{fig:howfs_ims}.  Note that in this case, we can find misalignment by looking at the relative lengths of the legs, as well as the relative centers of the legs (given by the intersection of linear fits to the leg pixels) and the apodizer center.  Because the apodizer blocks light in the center of the image, and we expect the poke pattern to be symmetric, differences in leg lengths indicate that the apodizer is off-center.

The same approach is also highly useful for tracking misalignment and rotations of the Lyot stop with the pupil viewer.  In this case, the lines being fit aren't bright regions generated by shapes put on the MEMS DM, but rather, dark areas generated by the features of the Lyot stop itself to block the telescope spiders.  \reffig{fig:lyot_spiders} illustrates the clustering procedure as applied in this case.  Once the pixels belonging to the four Lyot stop sections are clustered, it is trivial to find linear fits to the data by way of principal component analysis (PCA).  We have to use PCA rather than simple least-squares because we are fitting a 2D data set and must minimize errors in both directions, rather than just one direction as in a univariate least-squares fit.  As we are only interested in the line best fitting each set of pixels, we just need to find the eigenvector of the data covariance corresponding to the major eigenvalue.  For each set, we package the pixel coordinates into vectors $\mf x$ and $\mf y$ and form the augmented matrix:
\begin{equation}
R = \begin{bmatrix} \mf x - \mu_{\mf x} & \mf y - \mu_{\mf y} \end{bmatrix} \,,
\end{equation}
where $ \mu_{\mf x}$ and $\mu_{\mf y}$ are the sample means of $\mf x$ and $\mf y$, respectively, and $R$ has dimensionality $2 \times N$ for an $N$ pixel set.  The covariance of $R$ is thus given by:
\begin{equation}
S = \frac{R R^T}{N-1} \,,
\end{equation}
where $(\cdot)^T$ represents the transpose operator, and $S$ can be decomposed as:
\begin{equation}
S V = V \Lambda \,,
\end{equation}
where the columns of $V$ are the eigenvectors of $S$ and $\Lambda$ is the diagonal matrix of the corresponding eigenvalues.  Assuming that $\Lambda$ is ordered in decreasing magnitude, the first column of $V$ is the major mode of the data and represents the best linear fit to the set.  This method is both computationally efficient and find the minimum distance fit in both dimensions, so it is not biased by varying numbers of pixels at different locations in the set in either dimension, as seen in  \reffig{fig:lyot_spiders}.

This procedure, along with the one described in \S\ref{sec:ellipse_finding}, allows us to find position and rotation information to better than one tenth of one detector pixel in multiple pupil planes (as measured by the variance of successive fits), allowing us to meet our alignment requirements.  Furthermore, they require minimal image pre-processing (typically just a median filter and binary value conversion) and are widely applicable to any alignment task that involves imaging of regular features.

\section{Satellite Finding}\label{sec:satfinding}
As previously discussed, the grids printed on the apodizers generate satellite spots in the focal plane images.  These spots shift with pointing and centering changes, and the distance between spots scales with wavelength, but they should always form a perfect square pattern in the final image.  This property allows us to construct and algorithm that efficiently finds these spots without making any prior assumptions as to their location or separation, or requiring any additional knowledge other than the image contents. To begin, we note that given four points $\{\mf r_i\}_{i=1}^4$, we can define the set of six distance between them as:
\begin{equation}\label{eq:cond4}
\{ d_k\}_{k=1}^6 = \left\{\Vert \mf r_i - \mf r_j \Vert : i,j \in \{\nCk{4}{2}\} \right\} \,,
\end{equation}
where $ \{\nCk{4}{2}\}$ is the set of two element tuples representing all of the combinations of four elements taken two at a time, without repetition. Assuming the set is ordered by increasing magnitude, the four points define a square if and only if the first four distances are equal, the last two distances are equal, and the ratio of the magnitudes of the two subsets of distances is $\sqrt{2}$.

Of course, if we were to consider every pixel in an image, every single one of them would be a member of at least one such set.  Instead, we prioritize pixels by their magnitude, making the assumption that the satellite spots are relatively bright compared with the rest of the image (although not necessarily the brightest things in the image).  We construct a list of the brightest points in the image iteratively, identifying the current brightest point in the image, and then removing it, along with a radius of $x$ pixels about it, from the search region (the value of $x$ is determined by the average size of a satellite spot in all possible wavelengths).  These bright images are stored in the expanding set $\{\mf r_i\}_{i=1}^N$, where each vector is the 2D pixel coordinate of each bright spot.  Our task then becomes akin to a breadth-first search, where each node is itself a subtree whose branches represent all of the combinations of the root node with three other elements of  $\{\mf r_i\}$.  The value of all terminal nodes is binary---either the path to the node matches the condition on the distance set in \refeq{eq:cond4}, or it does not.  The search is terminated as soon as the condition is met.

Given this framework, it is clear that the cost of expanding each level of the tree grows with the depth---each additional point to consider increases the size of $\{\nCk{n}{4}\}$ by a factor of $(n+1)/(n-3)$.  Thus the search strategy described above is the lowest cost one, and preferable to first collecting all of the candidate points and then performing a depth-first search.  However, the algorithm is by no means optimal, as it will repeatedly deepen paths which cannot represent the vertices of a square.  We must therefore introduce pruning and some heuristic function to guide it. \cite{russell1992efficient}  We do so by noting that for any subset of three of the four vertices, the corresponding subset of of distances:
\begin{equation}\label{eq:cond3}
\{ d'_k\}_{k=1}^3 = \left\{\Vert \mf r_i - \mf r_j \Vert : i,j \in \{\nCk{3}{2}\} \right\} \,,
\end{equation}
must contain two elements that are equal, and one element that is $\sqrt{2}$ times larger.  This is a necessary, but not sufficient, condition for defining a square.  This condition allows us to prune whole branches of the tree, and greatly improve the search efficiency.  Note that this is the smallest subtree that can be tested, as any two points can form an edge of a square.

Operationally, the search is implemented as follows:  The set $\{\mf r_i\}_{i=1}^N$ is constructed iteratively as described above.  Once we have identified three bright spots, we begin maintaining a list of candidate groups that meet the condition on the distance set in \refeq{eq:cond3}.  These are stored in the subset $\{\mf r_{i\in C}\}$ where $C$ is the set of all candidate indices, of length $n_C$.  For each new spot after the first three, we construct the set of 3-element tuples, $ \{\nCk{n_C}{3}\} $ representing the combinations of elements of the candidate set taken three at a time.  Each of these tuples is augmented with the index of the newest point $N$ to create the set of 4-element tuple, $ \{G_i\}_{i=1}^{n_G}$, where
\begin{equation}
n_G = \frac{n_C!}{3!(n_C-3)!} \,.
\end{equation}
For each of these, we then construct the set of distances:
\begin{equation}\label{eq:distset}
\left(\{ d_k\}_{k=1}^6\right)_l = \left\{\Vert \mf r_i - \mf r_j \Vert : i,j \in G_l \right\} \,,
\end{equation}
and test whether it meets the condition for defining a square.  If the condition is not met, the candidate subset is updated in a similar fashion---testing the combinations of $ \{\nCk{N}{2}\}$ augmented with $N$ with the condition on the distance set in \refeq{eq:cond3}.  The next brightest point is then identified, and the next iteration started. 

\reffig{fig:satfinding1} illustrates the application of this algorithm to one slice of a GPI IFS image.  The satellite spots in this case are quite bright, but dimmer than much of the core of the image, and there is much bright speckle noise throughout the image.  The image is first convolved with a 2 dimensional Gaussian to remove hot and cold pixels and smooth out the image (essentially acting as a matched filter for the satellite spots).  The resultant image then has the algorithm applied to it.  The satellite spots are sequentially added as the fourth through seventh elements of the candidate spot set.  However, as all of the first three spots are clustered close together in the center of the image and do not form a right triangle, they do not meet the distance criteria associated with \refeq{eq:cond3} and are therefore not included in any tested combinations.  Once the candidate set has been found, each aperture surrounding the candidate locations is again run through a matched filter to find the exact sub-pixel location of the satellite spots' centers.  The algorithm works equally well in cases where the satellite spots are significantly dimmer, as demonstrated in \reffig{fig:satfinding2}.  In this case, the final satellite spot was only added to the candidate list on the 38th iteration, meaning that there were 73815 possible combinations of four spots, or 7770 possible combinations of three spots from the previous iteration.  However, only 28 subsets (0.36\%) had been identified as possible candidates for testing, so that the entire execution time was only 1.45 times longer than in the case shown in \reffig{fig:satfinding1}.

\section{Conclusions}
As coronagraphic systems reach higher contrasts they will become more and more sensitive to small environmental effects such as thermal variations and flexure, and will have progressively tighter tolerances on internal alignments.  This will require ever more components to become motorized (as continually realigning by hand will take too much time), which will allow for automation of all repeated alignment tasks.  Here,  we have demonstrated the utility of applying relatively simple computer vision techniques to perform just such tasks, using  both the engineering and science image data generated by the Gemini Planet Imager.  The routines described in this paper are currently in regular use as part of GPI integration and testing, and will continue to be employed as part of GPI's daytime calibrations when it goes on sky.  The motor settings generated in these calibrations will be used as set points for the various mechanical components, and will be updated by thermal models, which themselves have been verified via the same alignment tests.

These algorithms are all easy to implement and adapt to specific systems, and can greatly simplify any task requiring the repeated discovery of semi-regular features. As such tasks are quite common in astronomical imaging and instrumentation, there is a great potential for future applications of these methods, as well as multiple others commonly employed in computer vision.  In particular, the satellite spot finding algorithm described in \S\ref{sec:satfinding} is generally applicable to any problem in which one must find the vertices of regular geometric figures (i.e., any regular grid) in noisy image data.  In fact, the implementation written for finding the satellite spots was applied, unchanged, to the regular shape made by poking MEMS actuators (as in \reffig{fig:pupalign}) and automatically detected the generated spots in all instances. We fully expect that algorithms such as these will play an ever greater role in regular operations of future astronomical instrumentation and will be increasingly applied to automated data processing.

\section*{Acknowledgments}
The authors would like to thank the entire GPI collaboration, and especially the GPI integration and testing team for their support and dedication.  Portions of this work were performed under the auspices of the U.S. Department of Energy by Lawrence Livermore National Laboratory under Contract DE-AC52-07NA27344.  This paper is number LLNL-JRNL-599792.


\clearpage

\clearpage
\begin{table}[h]
\caption{\label{tbl:tolerances} GPI alignment tolerances}
\begin{center}
\begin{tabular}{cp{1in}c c}
\hline
Alignment & Tolerance$^a$\ & Measurement & Achieved$^b$\\ 
\hline
Telescope Pupil to MEMS &  0.5\% & AO WFS$^c$ & 0.012\%\\
AO WFS to MEMS &  0.5\%& AO WFS & 0.019\% \\
Apodizer to MEMS &  0.25\%& Pupil viewer$^d$ or HOWFS & 0.21\%\\
Lyot Stop to MEMS & 1\%& Pupil viewer & 0.48\%\\
\hline
\end{tabular}
\end{center}
\footnotesize $^a$Tolerances are given in percentages of the corresponding pupil size.\\
$^b$ Achieved values are RSS errors in tip and tilt on the WFS and errors in x and y on the pupil viewer.\\
$^c$ Measurements in the AO WFS are in fractions of sub-apertures, with 43 sub-apertures across the pupil.\\
$^d$ The pupil is approximately 232 pupil viewer pixels across.
\normalsize
\end{table}

\clearpage

\section*{List of Figure Captions}

\begin{figure}[ht]
 \begin{center}
   \includegraphics[width=0.95\textwidth]{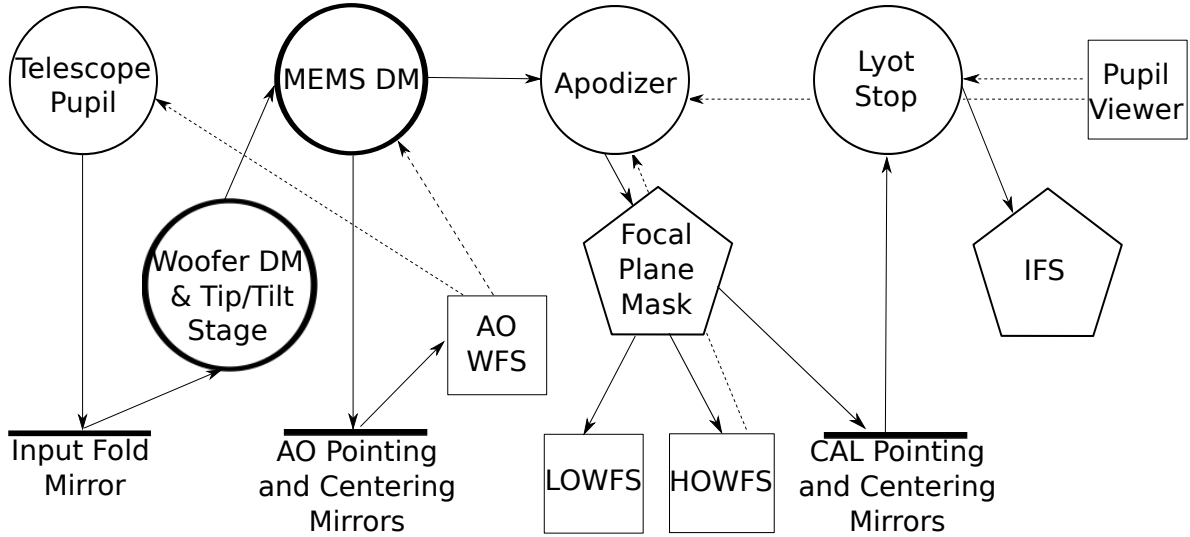} 
 \end{center}
 \caption[]{ \label{fig:gpi_schematic} Schematic of GPI.  Circles represent pupil planes, pentagons represent focal planes, and squares are wavefront sensors and detectors.  Thick lines represent controllable surfaces such as moveable mirrors and stages, lines with arrows represent the light path, and dashed lines represent imaging of pupil planes by various detectors for alignment purposes.  Note that this is a schematic view only: multiple optics, including beam splitters and pupil re-imagers, are not shown here.  The four pupil planes at the top of the schematic must all be aligned to a common plane, with tolerances given in Table \ref{tbl:tolerances}.}
 \end{figure}

\reffig{fig:gpi_schematic}  Schematic of GPI.  Circles represent pupil planes, pentagons represent focal planes, and squares are wavefront sensors and detectors.  Thick lines represent controllable surfaces such as moveable mirrors and stages, lines with arrows represent the light path, and dashed lines represent imaging of pupil planes by various detectors for alignment purposes.  Note that this is a schematic view only: multiple optics, including beam splitters and pupil re-imagers, are not shown here.  The four pupil planes at the top of the schematic must all be aligned to a common plane, with tolerances given in Table \ref{tbl:tolerances}.\\

\begin{figure}[ht]
 \begin{center}
   \includegraphics[width=0.5\textwidth]{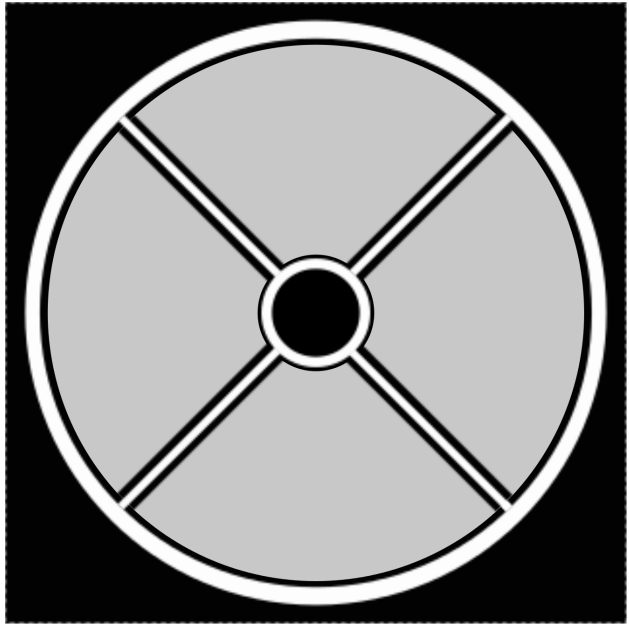} 
 \end{center}
 \caption[]{ \label{fig:pupil_schematic} Schematic of image seen at the GPI IFS pupil.  The black regions represent the Lyot stop, the gray regions are light that passes through the Lyot stop, and the white regions are light from the telescope pupil spiders  and the bright rings diffracted from the focal plane mask, which is stopped by the Lyot stop. This illumination pattern would be approximately what we would see if we could image the reverse side of the Lyot stop.  An actual image of the IFS pupil, with only the Lyot stop in the beam path, is shown in \reffig{fig:lyot_spiders}.  The apodizer, FPM and Lyot stop must all be closely aligned for the coronagraph to work properly. }
 \end{figure}

\begin{figure}[ht]
 \begin{center}
   \includegraphics[width=0.32\textwidth]{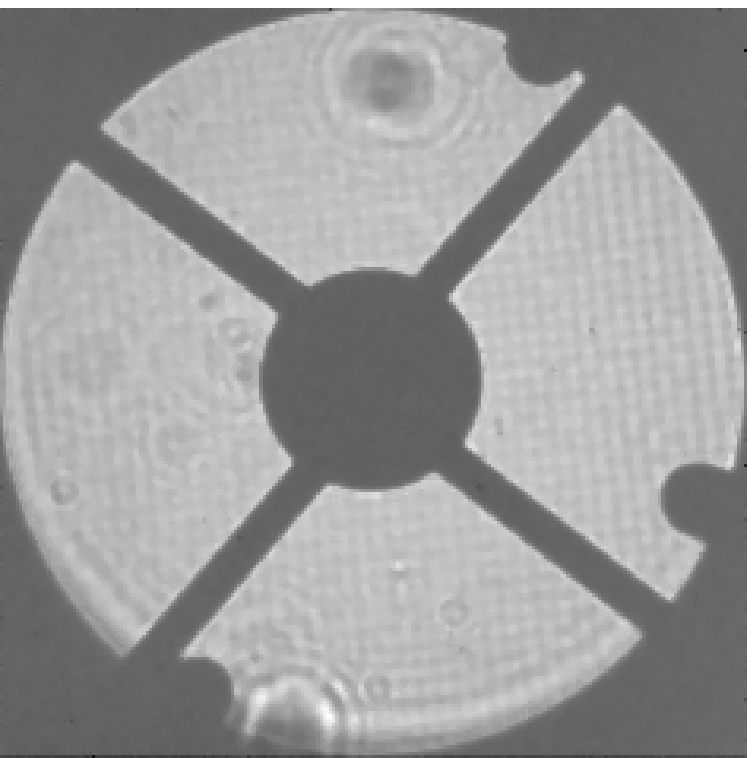} 
   \hspace{-2ex}
   \includegraphics[width=0.32\textwidth]{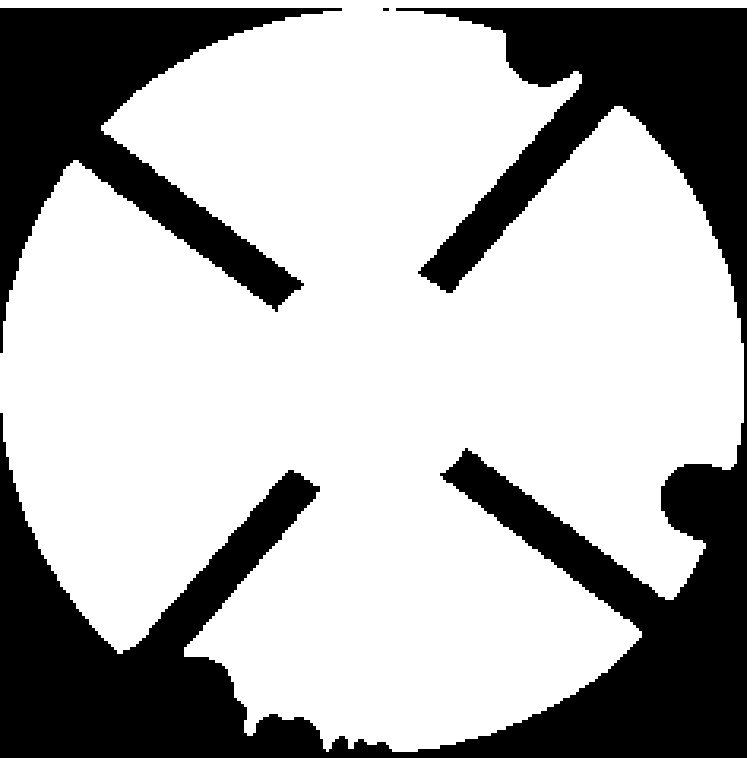}
     \hspace{-2ex}
   \includegraphics[width=0.35\textwidth,clip=true,trim=0.2in 0.1in 0.125in 0.25in]{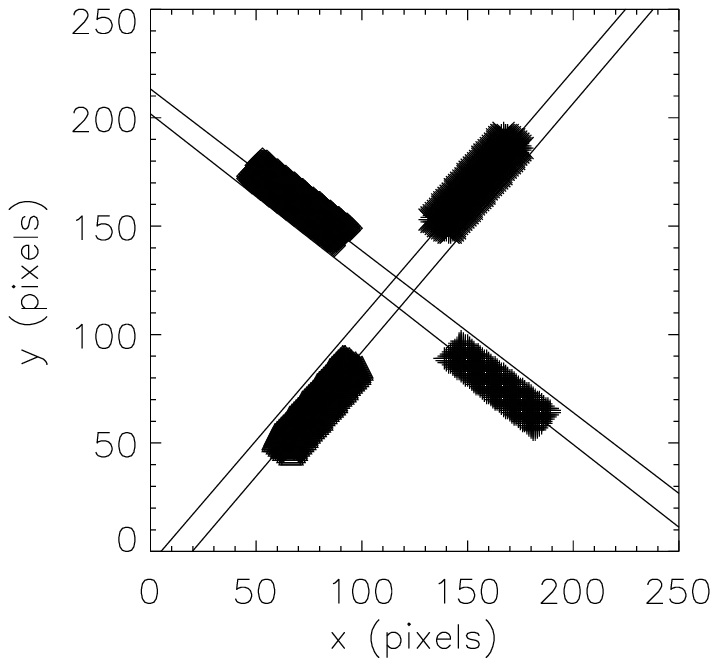}
 \end{center}
 \caption[]{ \label {fig:lyot_spiders}  Pupil viewer images of Lyot stop. \emph{Left}: Dark subtracted pupil viewer image, cropped to show Lyot stop only. \emph{Center}: The original image, median filtered and converted to binary units using the median image value.  The central obscuration of the stop is removed by using the algorithm from \S\ref{sec:ellipse_finding}.  The remaining features are the stop spiders and tabs used to block bad DM actuators.  The tabs at the edge of the pupil are removed by applying a single annulus scaled from the fit to the central obscuration.  \emph{Right}:  The four clusters of pixels belonging to the four spiders automatically detected and with best fit lines.}
 \end{figure}

\reffig{fig:pupil_schematic} Schematic of image seen at the GPI IFS pupil.  The black regions represent the Lyot stop, the gray regions are light that passes through the Lyot stop, and the white regions are light from the telescope pupil spiders  and the bright rings diffracted from the focal plane mask, which is stopped by the Lyot stop. This illumination pattern would be approximately what we would see if we could image the reverse side of the Lyot stop.  An actual image of the IFS pupil, with only the Lyot stop in the beam path, is shown in \reffig{fig:lyot_spiders}.  The apodizer, FPM and Lyot stop must all be closely aligned for the coronagraph to work properly.\\

\begin{figure}[ht]
 \begin{center}
   \includegraphics[width=0.33\textwidth]{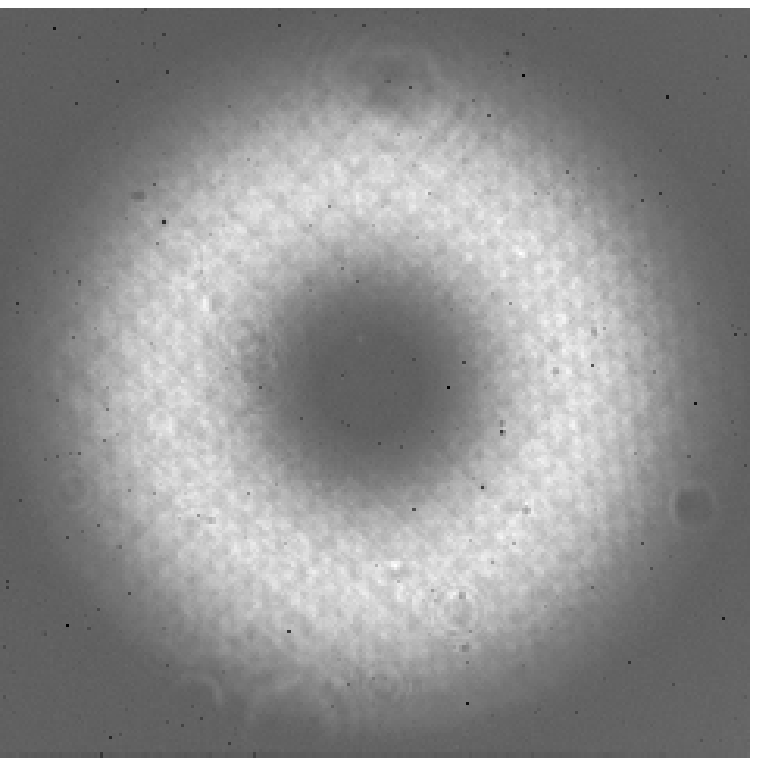} 
   \hspace{-2ex}
   \includegraphics[width=0.33\textwidth]{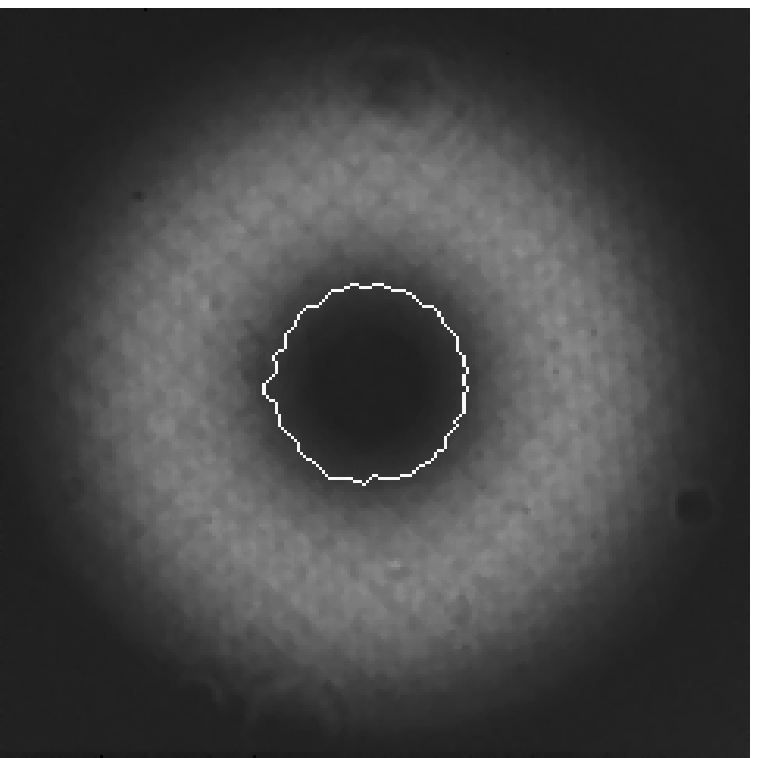}
     \hspace{-2ex}
   \includegraphics[width=0.33\textwidth]{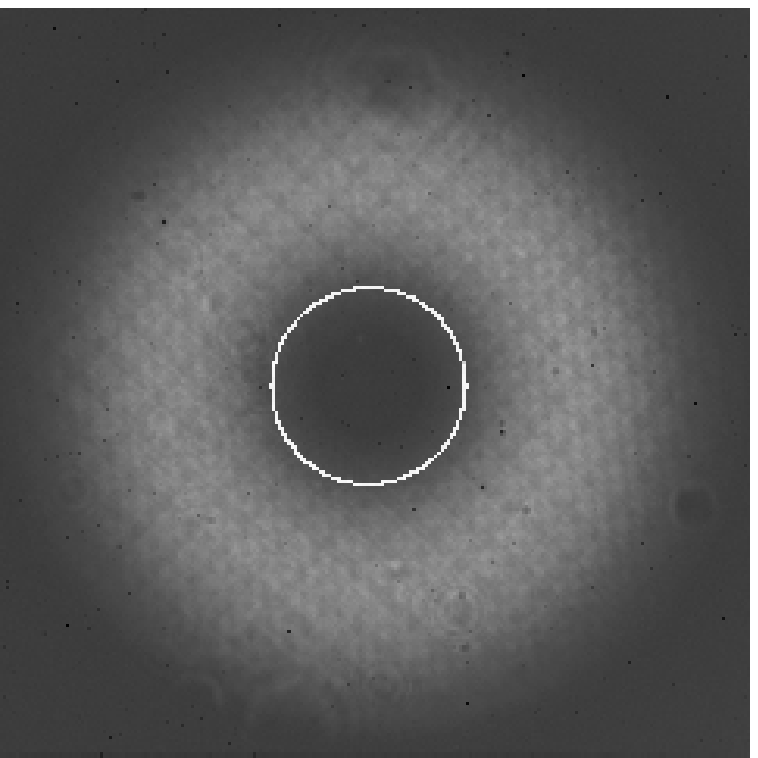}
 \end{center}
 \caption[]{ \label{fig:pupil_hough} Pupil viewer images of the apodizer only  (no FPM or Lyot stop in light path). \emph{Left}: Dark subtracted pupil viewer image, cropped to show the image of the apodizer only. \emph{Center}: Median filtered image with the candidate pixels for the inner annulus of the apodizer overlaid. Pixels are selected as candidates when they differ from their neighbors by more than 25\% of the image median in intensity.  A mask is applied so that pixels belonging to the outer annulus are not considered. \emph{Right}:  Original image with best solution from Hough transform overlaid. }
 \end{figure}

 \reffig{fig:pupil_hough} Pupil viewer images of the apodizer only  (no FPM or Lyot stop in light path). \emph{Left}: Dark subtracted pupil viewer image, cropped to show the image of the apodizer only.  \emph{Center}: Median filtered image with the candidate pixels for the inner annulus of the apodizer overlaid. Pixels are selected as candidates when they differ from their neighbors by more than 25\% of the image median in intensity. A mask is applied so that pixels belonging to the outer annulus are not considered. \emph{Right}:  Original image with best solution from Hough transform overlaid.\\
 
\begin{figure}[ht]
 \begin{center}
   \includegraphics[width=0.33\textwidth]{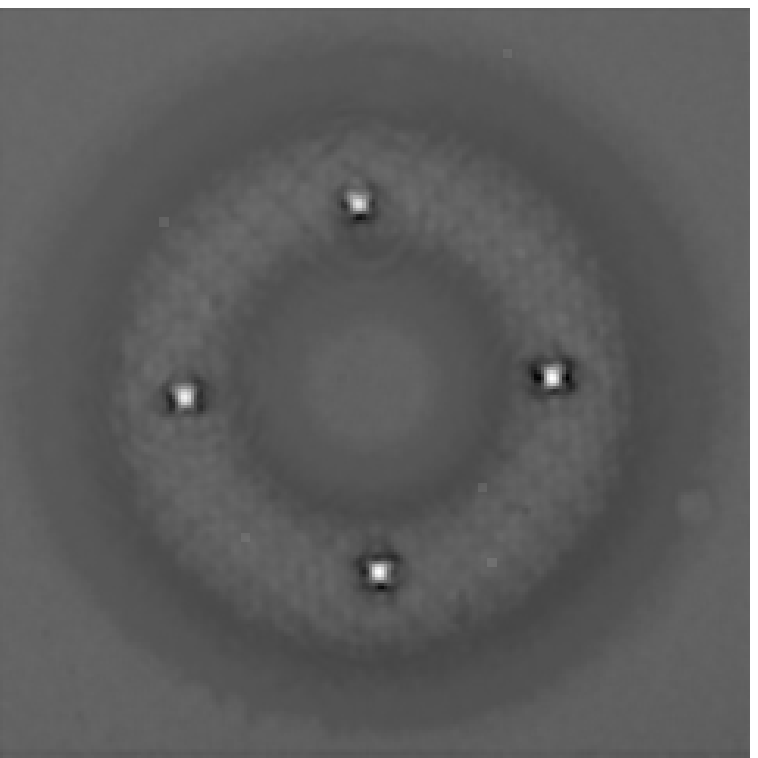} 
   \hspace{-2ex}
   \includegraphics[width=0.33\textwidth]{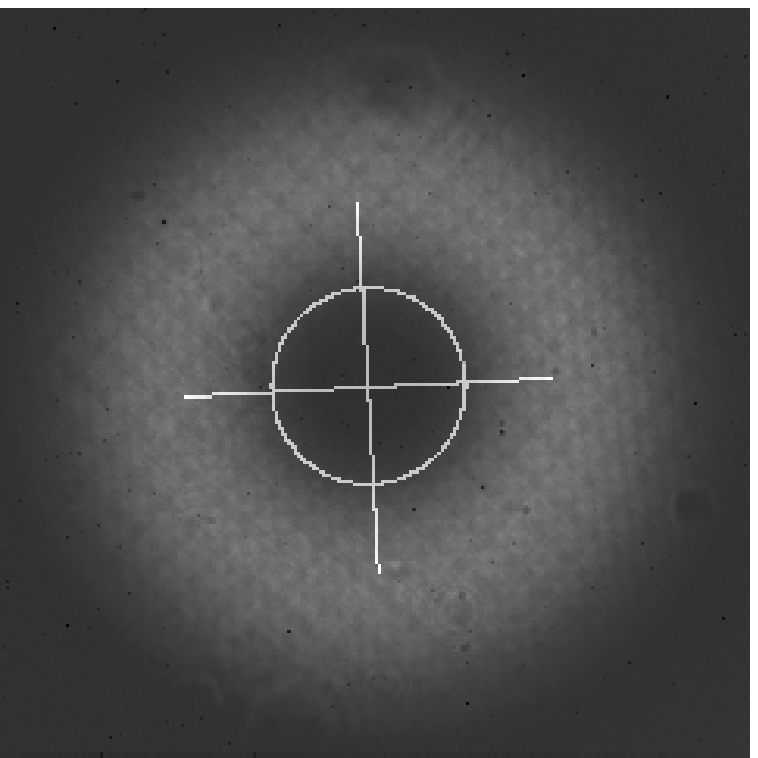}
 \end{center}
 \caption[]{ \label{fig:pupalign} \emph{Left}: Difference image between poked and flat DM states. \emph{Right}:  Original image with best solution from Hough transform and intersection of MEMS pokes overlaid. }
 \end{figure}

  \reffig{fig:pupalign} \emph{Left}: Difference image between poked and flat DM states. \emph{Right}:  Original image with best solution from Hough transform and intersection of MEMS pokes overlaid. \\
 
\begin{figure}[ht]
 \begin{center}
   \includegraphics[width=0.33\textwidth]{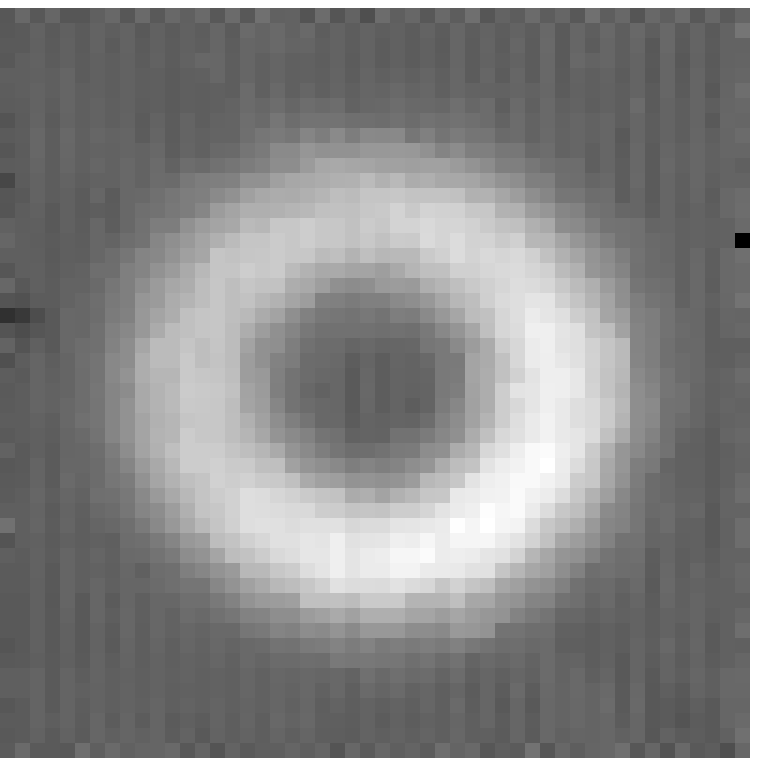} 
   \hspace{-2ex}
   \includegraphics[width=0.33\textwidth]{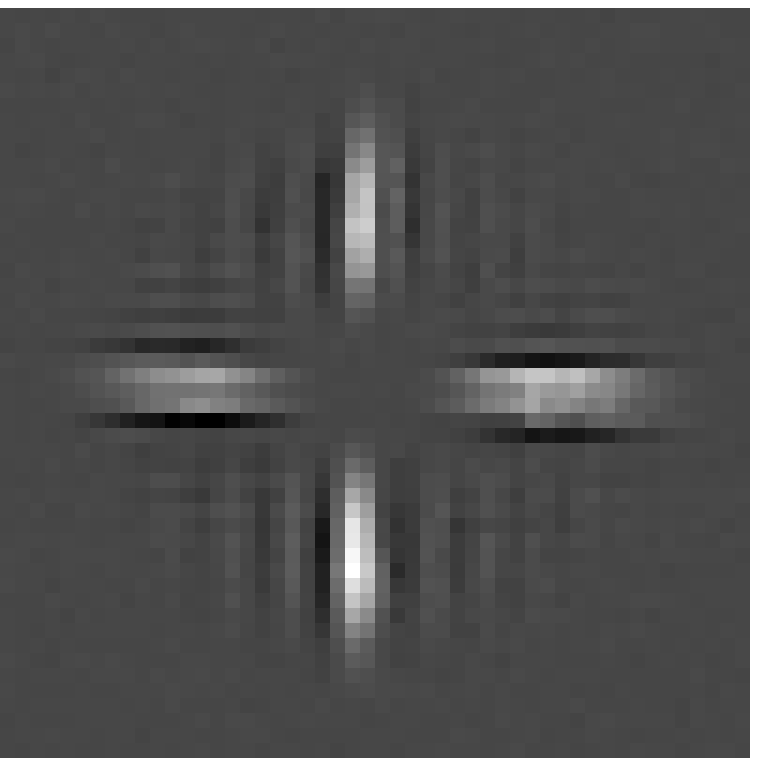}
     \hspace{-2ex}
   \includegraphics[width=0.33\textwidth,clip=true,trim=0.3in 0.1in 0.15in 0.25in]{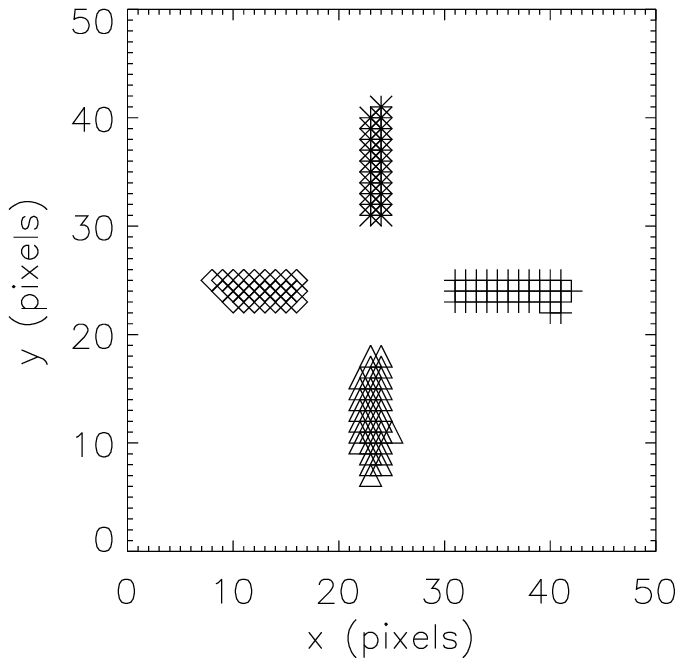}
 \end{center}
 \caption[]{ \label{fig:howfs_ims} HOWFS images of apodizer only (no FPM in light path). \emph{Left}: Median filtered HOWFS image, cropped to show the image of the apodizer only. \emph{Center}: Difference image between cross and flat DM shapes. \emph{Right}:  The four clusters of pixels automatically detected. }
 \end{figure}

  \reffig{fig:howfs_ims} HOWFS images of apodizer only (no FPM in light path). \emph{Left}: Median filtered HOWFS image, cropped to show the image of the apodizer only. \emph{Center}: Difference image between cross and flat DM shapes. \emph{Right}:  The four clusters of pixels automatically detected. \\
 
  \reffig {fig:lyot_spiders}  Pupil viewer images of Lyot stop. \emph{Left}: Dark subtracted pupil viewer image, cropped to show Lyot stop only. \emph{Center}: The original image, median filtered and converted to binary units using the median image value.  The central obscuration of the stop is removed by using the algorithm from \S\ref{sec:ellipse_finding}.  The remaining features are the stop spiders and tabs used to block bad DM actuators.  The latter are removed by applying a single annulus scaled from the fit to the central obscuration.  \emph{Right}:  The four clusters of pixels belonging to the four spiders automatically detected and with best fit lines.\\
 
\begin{figure}[ht]
 \begin{center}
   \includegraphics[width=0.33\textwidth]{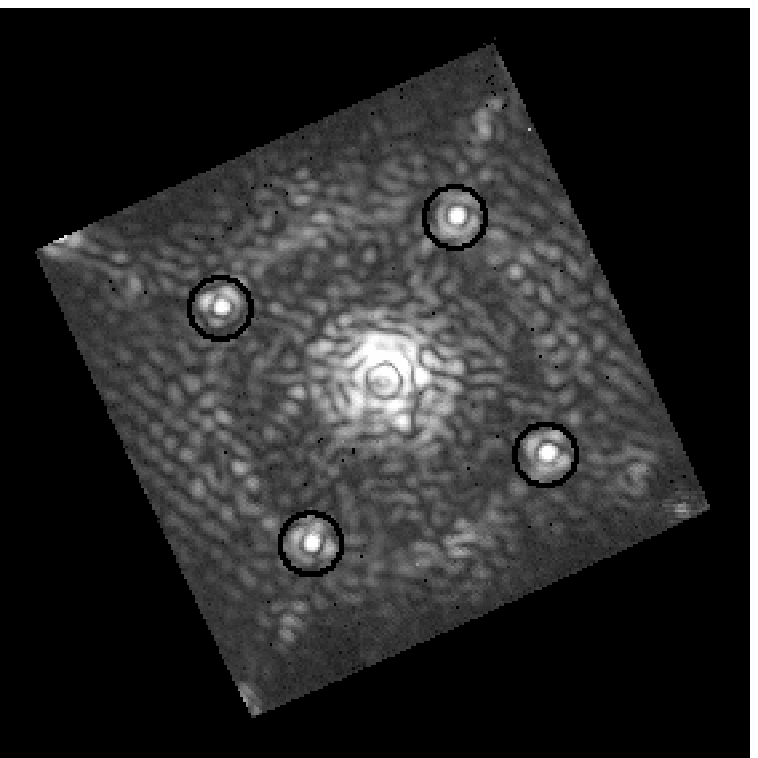} 
   \hspace{-2ex}
   \includegraphics[width=0.33\textwidth]{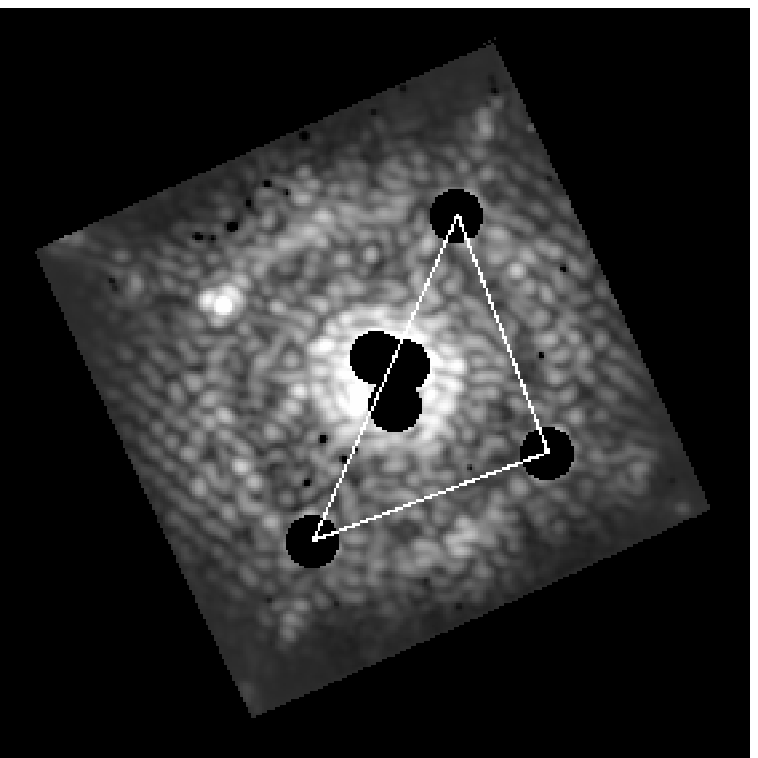}
     \hspace{-2ex}
   \includegraphics[width=0.33\textwidth]{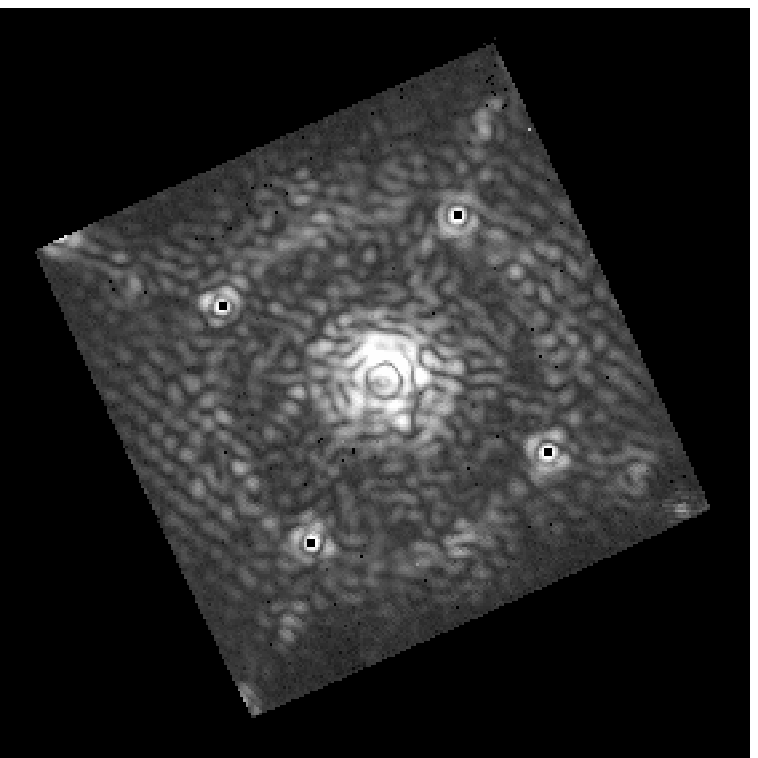}
 \end{center}
 \caption[]{ \label{fig:satfinding1} \emph{Left}: Single slice of reconstructed GPI IFS data cube.  The satellite spots are circled.  \emph{Center}: Penultimate step of satellite spot finding algorithm.  Although six spots have been identified as candidate spots, only three of them have the right geometry to be part of a square, and so the candidate subset only has one element, illustrated by the white triangle.  \emph{Right}:  The original image with the located satellite spots tagged with black dots. }
 \end{figure}

  \reffig{fig:satfinding1} \emph{Left}: Single slice of reconstructed GPI IFS data cube.  The satellite spots are circled.  \emph{Center}: Penultimate step of satellite spot finding algorithm.  Although six spots have been identified as candidate spots, only three of them have the right geometry to be part of a square, and so the candidate subset only has one element, illustrated by the white triangle.  \emph{Right}:  The original image with the located satellite spots tagged with black dots. \\
 
\begin{figure}[ht]
 \begin{center}
   \includegraphics[width=0.33\textwidth]{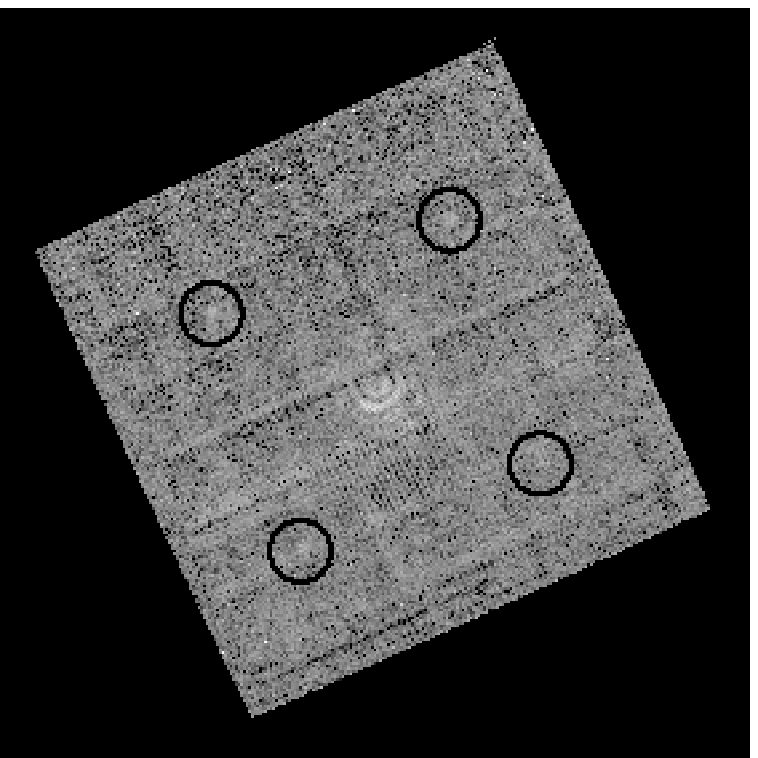} 
   \hspace{-2ex}
   \includegraphics[width=0.33\textwidth]{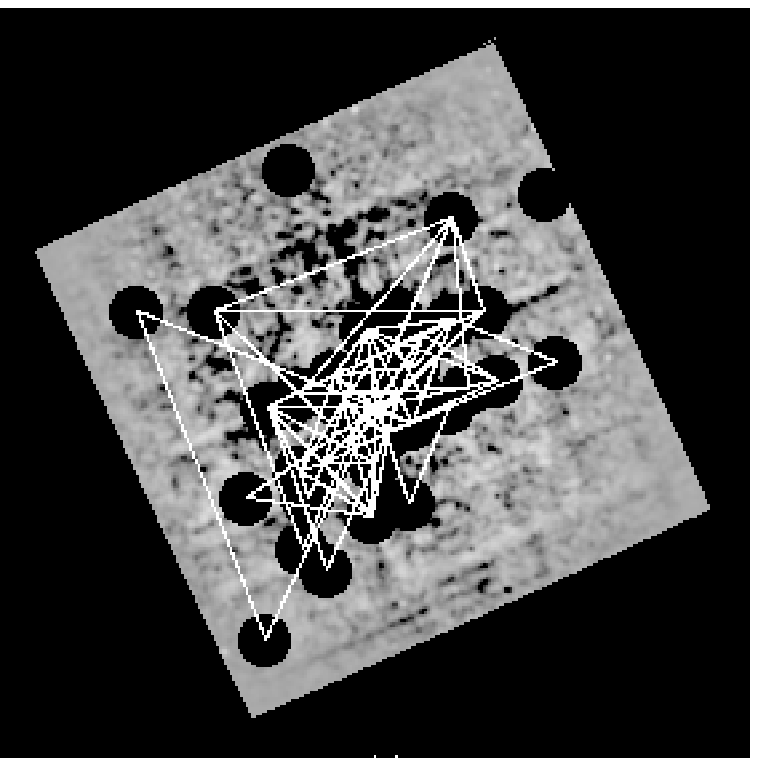}
     \hspace{-2ex}
   \includegraphics[width=0.33\textwidth]{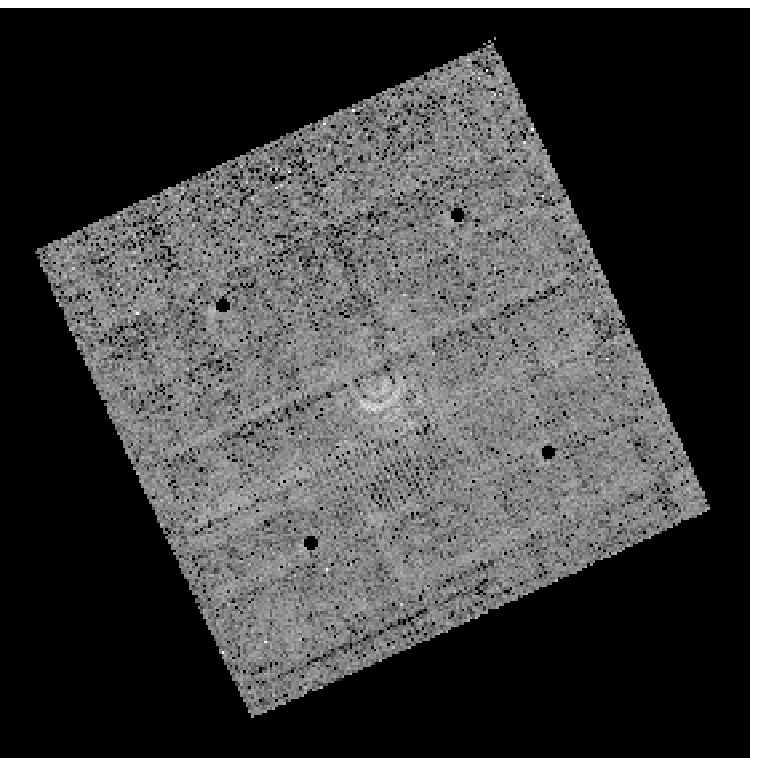}
 \end{center}
 \caption[]{ \label{fig:satfinding2} Results of satellite finding, arranged as in \reffig{fig:satfinding1}.  In this case, 38 candidate spots were identified before the satellite spots were found, but only 28 candidate subsets required testing at the final step of the iteration. The satellite spots were approximately 50 times fainter than in the previous example,  and the contrast in this image was one order of magnitude lower.}
 \end{figure}

 \reffig{fig:satfinding2} Results of satellite finding, arranged as in \reffig{fig:satfinding1}.  In this case, 38 candidate spots were identified before the satellite spots were found, but only 28 candidate subsets required testing at the final step of the iteration. \\

\clearpage

\clearpage

\clearpage

\clearpage

 \clearpage

 \clearpage

 \clearpage

 \clearpage

\end{document}